\documentclass[a4paper]{article}
\usepackage{lscape}
\usepackage[usenames,dvipsnames,svgnames,table]{xcolor}

\usepackage{tikz}
\usepackage{graphicx}
\usepackage{multirow}
\usepackage{amssymb}
\usepackage[right,modulo]{lineno}
\usepackage{array}
\newcolumntype{L}[1]{>{\raggedright\let\newline\\\arraybackslash\hspace{0pt}}m{#1}}
\newcolumntype{C}[1]{>{\centering\let\newline\\\arraybackslash\hspace{0pt}}m{#1}}
\newcolumntype{R}[1]{>{\raggedleft\let\newline\\\arraybackslash\hspace{0pt}}m{#1}}

\usepackage[top=7cm,bottom=2.5cm,headheight=120pt,headsep=15pt,left=6cm,right=1.5cm,marginparwidth=4.2cm,marginparsep=0.5cm]{geometry}

\usepackage{marginnote}
\reversemarginpar  
\usepackage{lipsum} 
\usepackage{calc}
\usepackage{siunitx}
\usepackage{lineno}
\usepackage[utf8]{inputenc}
\usepackage[T1]{fontenc}
\usepackage[
backend=biber,
natbib=true,
sortcites=true,
defernumbers=true,
style=authoryear,
citestyle=authoryear-comp,
maxnames=99,
maxcitenames=2,
uniquename=init,
giveninits=true,
terseinits=true,
url=false
]{biblatex}
\DeclareNameAlias{default}{family-given}
\renewbibmacro{in:}{%
  \ifentrytype{article}{}{\printtext{\bibstring{in}\intitlepunct}}} 
\DeclareFieldFormat[article]{pages}{#1} 
\AtEveryBibitem{\ifentrytype{article}{\clearfield{number}}{}} 
\DeclareFieldFormat[article, inbook, incollection, inproceedings, misc, thesis, unpublished]{title}{#1} 
\usepackage{csquotes}
\RequirePackage[english]{babel} 
\DeclareBibliographyCategory{ignore}
\addtocategory{ignore}{recommendation} 
\usepackage{nameref}
\usepackage[pdfborder={0 0 0}]{hyperref}  
\usepackage{graphbox}  
\usepackage{microtype}
\DisableLigatures[f]{encoding = *, family = * }
\RequirePackage[default,scale=0.90]{opensans}
\usepackage{xcolor}
\definecolor{darkgray}{HTML}{808080}
\definecolor{mediumgray}{HTML}{6D6E70}
\definecolor{ligthgray}{HTML}{d9d9d9}
\definecolor{pciblue}{HTML}{74adca}
\definecolor{opengreen}{HTML}{77933c}
\usepackage{changepage}
\usepackage[aboveskip=1pt,labelfont=bf,labelsep=period,singlelinecheck=off]{caption}

\usepackage{fancyhdr}  
\usepackage{lastpage}  
\fancyhfoffset[L]{4.5cm}  
\renewcommand{\headrulewidth}{\ifnum\thepage=1 0.5pt \else 0pt \fi} 

\newcommand{\PCI}{Peer Community In Evolutionary Biology}

\newcommand{\beginingpreprint}{
\vspace*{0.5cm}
\begin{flushleft}
\baselineskip=30pt
\marginpar{
\large\textnormal{\color{pciblue}\\RESEARCH ARTICLE}\\
\vspace*{0.5pt}
\\
\includegraphics[align=c,width=0.5cm]{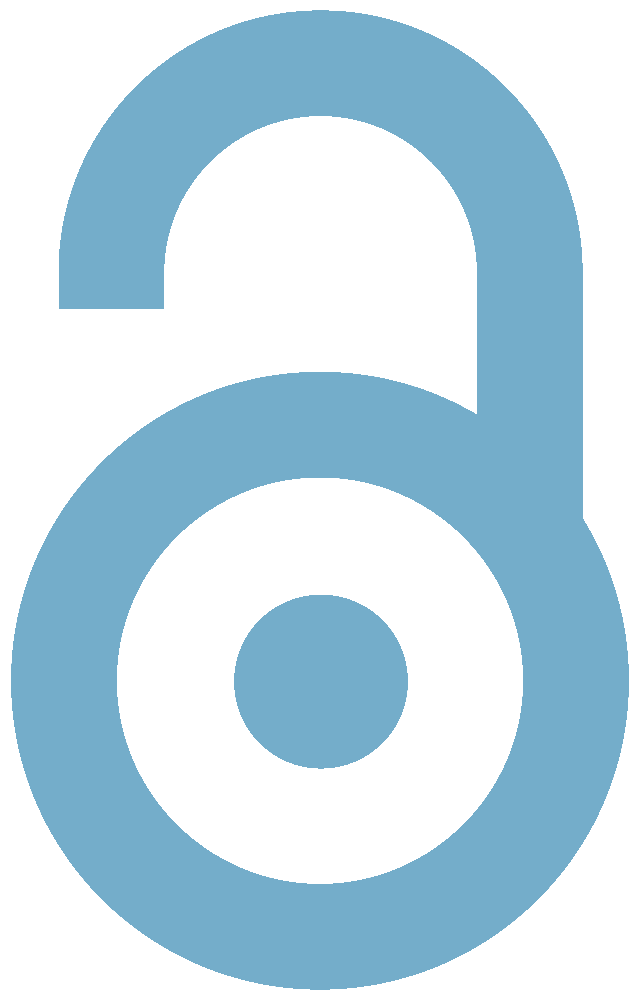} \space \large\textbf{\color{pciblue}Open Access}\\
\\
\includegraphics[align=c,width=0.5cm]{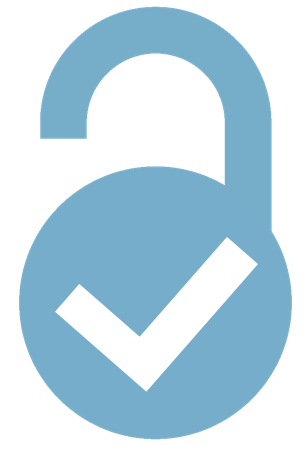} \space \large\textbf{\color{pciblue}Open Peer-Review}\\
\\
\\
\\
\\
\raggedright
\scriptsize\textbf{Cite as:}\space
Pouyet F and KJ Gilbert (2020). Towards an improved understanding of molecular evolution: the relative roles of selection, drift, and everything in between. \textit{arXiv} 1909.11490 [q-bio], ver.4 peer-reviewed and recommended by \textit{PCI Evolutionary Biology}. url:~\url{https://arxiv.org/abs/1909.11490}\\
\vspace*{0.5cm}
\textbf{Posted:} \datepub\\
\vspace*{0.5cm}
\textbf{Recommender:}\\
\recommender\\
\vspace*{0.5cm}
\textbf{Reviewers:}\\
\reviewers\\
\vspace*{0.5cm}
\textbf{Correspondence:}\\
\href{mailto:\email}{\email}\\

}
{\Huge
\fontseries{sb}\selectfont{\preprinttitle}}
\end{flushleft}
\vspace*{0.25cm}
\begin{flushleft}

\Large
\listauthors
\end{flushleft}
\bigskip
{\raggedright
\listinstitutions}
\begin{flushleft}
\fcolorbox{lightgray}{lightgray}{
\parbox{\textwidth - 2\fboxsep}{
\centering\large{\fontseries{sb}\selectfont{This article has been peer-reviewed and recommended by\\
\emph{\PCI}}}\\
}}
\end{flushleft}
\vspace*{0.5cm}
\fcolorbox{pciblue}{pciblue}{
\parbox{\textwidth - 2\fboxsep}{
\vspace{0.25cm}
\textbf{\large{\textsc{Abstract}}}\\
\preprintabstract\\

\footnotesize{\textbf{\emph{Keywords: }}\preprintkeywords}
\vspace{0.25cm}}
}
}

\newcommand{\preprinttitle}{Towards an improved understanding of molecular evolution: the relative roles of selection, drift, and everything in between}

\newcommand{\listauthors}{\raggedright 
Fanny Pouyet\textsuperscript{1,2}, \&
Kimberly J. Gilbert\textsuperscript{1,3}
}

\newcommand{\listinstitutions}{
\textsuperscript{1} Institute of Ecology and Evolution, University of Bern -- Bern, Switzerland; Swiss Institute of Bioinformatics, Lausanne, Switzerland;
\\
\textsuperscript{2} Sorbonne Universite, CNRS, Institut de Biologie Paris-Seine, Laboratory of Computational and Quantitative Biology, 7-9 Quai Saint Bernard, 75005 -- Paris, France;\\
\textsuperscript{3} Department of Computational Biology, University of Lausanne (UNIL), Genopode 2016, 1015 -- Lausanne, Switzerland

}

\newcommand{\datepub}{18th June 2020}

\newcommand{\recommender}{Guillaume Achaz}

\newcommand{\reviewers}{Benoit Nabholz and one anonymous reviewer}

\newcommand{\email}{fanny.pouyet@gmail.com; kjgilbert.evolution@gmail.com}

\newcommand{\preprintabstract}{A major goal of molecular evolutionary biology is to identify loci or regions of the genome under selection versus those evolving in a neutral manner. Correct identification allows accurate inference of the evolutionary process and thus comprehension of historical and contemporary processes driving phenotypic change and adaptation. A fundamental difficulty lies in distinguishing sites targeted by selection from both sites linked to these targets and sites fully independent of selection. These three categories of sites necessitate attention in light of the debate over the relative importance of selection versus neutrality and the neutral theory. Modern genomic insights have proved that complex processes such as linkage, demography, and biased gene conversion complicate our understanding of the role of neutral versus selective processes in evolution. In this perspective, we first highlight the importance of the genomic and (a)biotic context of new mutations to identify the targets of natural selection. We then present mechanisms that may constrain the evolution of genomes and bias the inference of selection. We discuss these mechanisms within the two critical levels that they occur: the population level and the molecular level. We highlight that they should be taken into account to correctly distinguish sites across the genome subject to selective or non-selective forces and stress that a major current field-wide goal is to quantify the absolute importance of these mechanisms.}

\newcommand{\preprintkeywords}{natural selection; genetic drift; non-adaptive evolution; linkage}

\begin{document}
\beginingpreprint


\section*{Introduction}
\label{Intro}
Understanding the relative importance of evolutionary forces in driving adaptive change has been a longstanding goal of evolutionary biology. In today's genomic era, accurately and precisely addressing this question has become more feasible than ever before. Genomic data has allowed, for example, quantification of introgression rates between populations or species \parencite[e.g.][]{ellstrand2013introgression} and accurate estimation of mutation rates within species or across the genome \parencite{Hodginkson,ZhuMutRate,Besenbacher,ellegren2003mutation}. Yet the interplay between neutral evolution and selective forces has remained a difficult problem to address. 
Since the advent of population genetics as a field, debate over the relative importance of these processes has arisen, been resolved, and re-arisen  \parencite[e.g.][]{kimura1968evolutionary,ohta1971constancy,kreitman1996neutral,gillespie1995ohta}. Most recently, 50 years since the advent of the neutral theory, this debate has been rekindled in light of emerging genomic data \parencite{kern2018, jensen2018}. In an era of limited genetic tools and data, the neutral theory aimed to explain the greater than expected genetic diversity observed based on the actions of natural selection alone. Kern and Hahn (2018) have most recently argued that modern genomic data allows us to reject the applicability of neutral theory for understanding molecular evolution, while Jensen et al. (2018) have replied that this is not the case. 
A major dividing view on this point is whether a large proportion of the genome is affected by adaptive natural selection (directly or indirectly), and there is ample space for additional data across a wider range of species to contribute towards these investigations and our understanding of molecular evolution.

Natural selection functions in a diversity of modes. Negative selection -- also termed purifying selection -- acts to reduce the frequency of deleterious mutations (\emph{i.e.} mutations that reduce an individual's fitness, with selection coefficient $s<0$) while positive selection favors the fixation of beneficial mutations ($s>0$). Both these modes of selection reduce diversity by favoring or disfavoring specific alleles, but selection can also maintain genetic diversity when there is a selective advantage to being in the heterozygous state, \emph{e.g.} balancing selection. There is also the case of sexual selection that we do not address here as its many and complex cases merit a review of their own. 

Genetic drift is the neutral corollary to natural selection, where allele frequencies change due to random chance and sampling effects. Sites that are truly neutral are defined as those with $s=0$. If we define the effective size of a population as the size of an ideal population that experiences the same amount of genetic drift as the observed population \parencite{wright1931}, then the appropriate measure of the strength of selection is ($Ns$). This is key because genetic drift can act across a range of weakly selected sites when $N$ is sufficiently large, while at a smaller $N$, these sites may behave in a neutral manner.

Inference methods making use of empirical genomic data often require data solely originating from neutral processes, e.g. to infer demography with SFS-based methods such as FastSimCoal \parencite{fastimcoal}, or to infer the distribution of fitness effects (DFE) which requires constrasting SFS from neutral versus selected sites \parencite{keightley2007joint, tataru2019polydfev2}. 
\textit{A priori} misidentification of selected versus neutral sites may strongly bias resultant inferences, having a major impact on downstream interpretations. Approaches that search for signatures of selection and identify causal variants for adaptation or other phenotypic change mainly rely on identifying outlier regions of genomic differentiation \parencite[e.g GWAS, $F_{ST}$ outlier tests, or environment-allele correlations;][]{beaumont1996evaluating, whitlock2015reliable, luu2017pcadapt, foll2008approximate}, sometimes incorporating the signature across a stretch of the genome \parencite[e.g.][]{shic}.

Yet, a departure from genetic drift alone is not sufficient to merit a conclusion of selection. Even supervised machine learning methods that use summary statistics to infer a history of selective sweeps, such as S/HiC \parencite{shic}, are sensitive to confounding factors such as complex demographic history to accurately identify the variants under selection. 
In this review we highlight both the importance of considering the genomic, biotic, and abiotic context in which new mutations occur and the major evolutionary processes that can change allele frequencies, creating a major confounding factor for evolutionary inference of natural selection.

\section*{Genomic and environmental context}
The context in which mutations occur plays an important role in the actions of selection versus drift. This context encapsulates both the genomic environment as well as the biotic and abiotic environment of an organism containing those genes, therefore becoming relevant at both the molecular and population levels of interaction. When interactions between loci occurring together in the genome create non-additive phenotypic changes \parencite{fisher1918, cordell, martin} this can greatly complicate the inference of selection. In the presence of epistatic interactions, an allele at a given site may only be beneficial to an organism when its genomic environment contains another mutant allele at a different site in the genome, so that when together these alleles generate a phenotypic change. The concept of epistasis is tightly linked to the relative fitness effects of alleles, where a specific allele at one locus might change the sign of the selection coefficient at another locus. This phenomenon may create strong patterns of linkage disequilibrium and interfere with detection of selection since the selective effect is dependent on the combinations of alleles across loci. Such an example of epistasis is provided in the Segregation distortion paragraph.

In some cases, interference between selected sites of differential fitness effects can alter the strength of selection on a genomic region. More positively (or negatively) selected sites with physical linkage between each other can behave as a larger multi-site locus with an amplified selection coefficient representative of all the selected alleles in the region. 
In other cases, the strength of selection may be reduced when sites have competing impacts on fitness, termed Hill-Robertson interference \parencite{Felsenstein1974}. Amplification of the strength of selection by tightly-linked, jointly selected sites may simplify detection of selection but complicate the identification of precise sites under selection. Conversely, Hill-Robertson interference may complicate identifying both the presence of selection and the sites it targets. All of these effects depend on many parameters, for instance, with a single population undergoing partial self-fertilization, selective interference on deleterious alleles tends to reduce mean fitness and increase inbreeding depression. This effect is stronger when deleterious alleles are more recessive but only weakly dependent on the strength of selection against deleterious alleles and the recombination rate \parencite{Roze745}. Selective interference thus affects the relative impact of adaptive and non-adaptive processes in the genome.

Equally complicating is the individual-level scenario where fitness is dependent on the local community of organisms and whatever traits they exhibit (e.g. frequency-dependent selection) or on the abiotic local conditions exerted by variable environments on the phenotype (spatially or temporally varying selection). For example, in scale-eating cichlids, frequency-dependent selection can drive the handedness of individual phenotypes, where it is advantageous to be the rarer morph \parencite{hori1993frequency}. Interestingly, competition between alleles at the genomic level can also lead to frequency-dependent selection, with the most famous example being the amino acid polymorphism at position 6 of $\beta$-globin in Africa, which is associated with resistance to malaria \parencite[see][for review and meta-analyses]{TAYLOR2012}. Spatially varying selection results from environmental variation across geographic space \parencite[e.g.][]{Gagnaire725}. When populations exist across variable environments, distinct combination of alleles may arise within subpopulations locally adapted to their environments. Thus, the strength of selection on non-neutral sites can vary over time and space as an organism's environment changes. For the purposes of evolutionary inference and understanding the action of selection, experimental design is key so that empirical analyses can be conducted where the time point is equivalent and the environment is equivalent (or otherwise controlled for) to as much of an extent as possible \parencite[e.g.][]{Gorter}. Spatially and temporally varying selection may be best accounted for by this direct approach or by studying natural clines of allele frequencies \parencite{endler,machado}. However, for sites in the genome that are not targets of selection, nor purely under genetic drift, remaining processes can change allele frequencies in ways indicative of selection and are thus essential to bear in mind for evolutionary inference. 

In the remainder of this perspective, we consider evolutionary processes at both the population and molecular levels that have the potential to bias the inference of selection. Without consideration of such processes, inaccurate conclusions may be drawn about the role of selection in the evolutionary process. An extreme example of such mis-inference is Evans et al. \parencite*{Evans1717}
and Mekel-Bobrov et al. \parencite*{Mekel-Bobrov1720}
, where the result that selection drove brain size in humans was shown to be equivalently explained by neutral demographic processes \parencite{currat2006}. Scenarios involved at the molecular level can also interfere with correct inference of selection, including transmission bias \parencite[e.g. meiotic driver genes; ][for a review]{meioticdrive}, biased gene conversion \parencite[e.g. human accelerated genomic regions, HAR;][]{pollard, galtier2007}, or inference of the DFE \parencite[e.g. in flycatchers;][]{bolivar}.

Though these are few examples, we argue that these population and molecular processes are prevalent enough to act as major determinants of genomic diversity and are particularly easy to confound with selection. We also argue that even if some cases of demographic processes may be versions of genetic drift, in that they are due to sampling process, the signatures that they leave in populations differs from that of single, ideal populations undergoing drift, emphasizing their importance in distinguishing selection from neutrality. We define these as non-adaptive processes since they are neither the direct action of selection nor are they purely subject to drift (Table \ref{table:summary}). This non-adaptive category should help to improve studies of both neutral or selective processes impacting the genome by considering the relevant sites impacted by either process.
\begin{landscape}

\begin{table}
\begin{tabular}{ll|l|l|l}
 &                                                                      & \begin{tabular}[c]{@{}l@{}}Acts on \\ neutral\\ variants?\end{tabular} & 
 \begin{tabular}[c]{@{}l@{}}Non-\\ adaptive?\end{tabular} & Explanation \\ \hline
Selection & Negative & & & \multirow{4}{*}{\begin{tabular}[c]{@{}l@{}}Adaptive processes \\  ~  increase fitness\end{tabular}} \\
 & Positive & &  & \\
 & Balancing & &  & \\
 & Sexual & &  & \\ \hline
 \multicolumn{2}{l|}{\begin{tabular}[c]{@{}l@{}}Genetic drift\end{tabular}} & yes & yes & \begin{tabular}[c]{@{}l@{}}Neutral and non-adaptive as independent of fitness; \\  ~ depending on $N$, may impact site(s) with $s \neq 0$\end{tabular} \\ \hline
\multicolumn{2}{l|}{\textbf{Population-level processes}}& & &\\
 \multicolumn{2}{r|}{Bottleneck, demographic expansion} & yes & yes & \begin{tabular}[c]{@{}l@{}} Lead to rapid genetic drift, can mimic positive \\ ~ or negative selection\end{tabular} \\
 \multicolumn{2}{r|}{ Population structure } & yes  & yes & \begin{tabular}[c]{@{}l@{}} Can mimic local adaptation, responsible for many \\  ~ GWAS false-positives \end{tabular}\\
 \multicolumn{2}{r|}{Spatial expansion (gene surfing) } & yes & yes & Strong genetic drift over space, can mimic selection \\ \hline

\multicolumn{2}{l|}{\textbf{Molecular-level processes}}& & &\\

\multirow{3}{*}{\begin{tabular}[c]{@{}r@{}}Linked\\ selection\end{tabular}} &
\begin{tabular}[c]{@{}c@{}} ~ Background \\ ~ ~ selection\end{tabular} & yes & yes & \multirow{3}{*}{\begin{tabular}[c]{@{}l@{}}Impact neutral diversity, can change \\ ~  fitness, and act on selected sites depending on\\  ~ the relative selection coefficients of nearby sites\end{tabular}} \\ 
 & \begin{tabular}[c]{@{}c@{}}Selective sweeps\end{tabular} & yes & yes &  \\
 & \begin{tabular}[c]{@{}c@{}}Associative \\ overdominance\end{tabular} &  yes & yes & \\ 
\multicolumn{2}{l|}{\begin{tabular}[c]{@{}l@{}}CpG hyper-mutability\end{tabular}} &  yes & yes & \begin{tabular}[c]{@{}l@{}}Methylated Cytosines deaminate to Thymines, increasing \\  ~  the frequency of T within the genome\end{tabular}\\ 
\multicolumn{2}{l|}{\begin{tabular}[c]{@{}l@{}}Segregation distortion\end{tabular}} &  yes & yes & \begin{tabular}[c]{@{}l@{}}Meiotic gene drivers manipulate their transmission \\  ~  even though they can be detrimental\end{tabular}\\ 

\multicolumn{2}{l|}{\begin{tabular}[c]{@{}l@{}}Gene conversion\end{tabular}} & yes & yes & \begin{tabular}[c]{@{}l@{}}Increases the frequency of certain variants\\ ~ deterministically; regardless of fitness\end{tabular}\\

\end{tabular}
\caption{\label{table:summary} Evolutionary processes discussed in this paper that impact the fate of genetic variants. We focus on SNPs in this table and omit a category on the mobility of genetic elements, but transposable elements are also impacted by molecular-level genomic linkage processes.}
\end{table}
\end{landscape}

\section*{Population-level processes}
\subsection*{Population size}
 A major difficulty in distinguishing neutral sites from those impacted by selection is demographic history. The term demography incorporates several factors, but at its core is defined as change in population size, $N$. Population size combined with the selection coefficient determines the effective strength of selection on existing genomic diversity and therefore has major effects on the evolutionary process. Changes in population size result during population bottlenecks, population expansions (within one locale), spatial population expansions over geographic space (e.g. range expansion), or migration among populations (a more complicated case where a larger gene pool becomes relevant). 
 
 While selection or drift may act on specific variants or regions of the genome, demographic change affects the whole genome equally. Population bottlenecks have long been known to impact genetic diversity and change the efficiency of selection acting on alleles with $s \neq 0$ within the population. In such a case, stronger drift impacts sites of both $s=0$ and $s$ close to zero (nearly neutral mutations), with an increasing range of $s$ values as population size decreases and selection becomes less efficient. These alleles are driven to more rapid fixation or more rapid loss, a pattern of allele frequency change which can mimic that of positive selection and selective sweeps (see the Linked selection paragraph).
 
 Inferring past population size bottlenecks is a rich field with many methods to do so from genomic data \parencite{liu2015exploring, heled2008bayesian, terhorst2017robust, li2011inference}. Importantly, these methods rely on the use of neutral variants to obtain a proper inference \parencite{gattepaille2013inferring}, and are perhaps particularly important in conservation genetics to identify species at risk due to a recent bottleneck \parencite[rather than, e.g. an incorrect inference suggesting low diversity is due to a selective sweep;][]{peery2012reliability}.
 The bias that results from a demographic history of range expansion after a bottleneck has been particularly notable for humans having expanded out of Africa and more troublesome for distinguishing demography from selection. Studies attempting to find signatures of selection in humans may suffer from biased inferences due to these neutral historic processes \parencite{heller2013confounding, amos2011using, martin2017human}. Such a demographic history is particularly intriguing as it combines not only impactful changes in population size, but movement over geographic space which includes complications of population structure and spatially varying selection. 
 
\subsection*{Population structure}
Spatial population genetic structure and migration among subpopulations also plays an important role in the inference of selection. Population substructure can mimic a signal of local adaptation, where some populations which happen to exist in different environments possess different genetic signatures, leading methods to identify these differentiated loci as targets of selection. For instance, several recent studies have encountered this difficulty where signals previously thought to be selective were instead due to the lack of accounting for genetic structure among populations \parencite{berg2019reduced, tian2008accounting}. Additionally, the process of migration into populations or admixture among species can create an influx of novel genetic material. Even if fully neutral, the presence of such heterozygosity in the population leaves a signal indicative of either adaptive processes (e.g. balancing selection) or non-adaptive processes (e.g. secondary contact or gene flow among structured populations) \parencite{hahn2019molecular}.

It is again of vital importance for studies inferring selection that population structure be identified and accounted for. Many such approaches exist and vary depending on the form in which structure presents itself: isolation by distance versus more distinct populations over space, with varying degrees of migration occurring across the landscape. For example, isolation-with-migration models aim to infer the amount of migration between isolated populations leading to the level of polymorphism observed \parencite{hey2010isolation}. Isolation by distance can also be difficult in the face of inferring selection due to the correlation of allele frequency changes over space with environmental changes. Fortunately, much work has been done to correct for this population structure when inferring selection over landscapes \parencite{gunther2013robust,caye2019lfmm,gautier2015genome,de2015new}

\subsection*{Spatial and temporal variation}
Spread or growth of populations across geographic space also introduces the complexity of changing environmental conditions (or analogously temporally changing environments may have similar impacts). Many populations and species are known to have undergone or are expected to undergo this demographic change: from post-glacial recolonizations, to species invasions, to shifting species ranges in response to climate change \parencite{davis2001range, thomas2010climate}. During spatial expansions, not only does population size change with repeated bottlenecks of founder individuals, but these populations colonize new geographic space, resulting in a process termed gene surfing. Gene surfing is a unique genetic process that can leave genomic signatures similar to those of selection, yet are due entirely to demographic processes. Sequential founder events reduce the effective population size in colonizing populations and thus the efficiency of selection, thereby allowing alleles that might otherwise be subject to strong selection to surf to high frequency at the expanding wave front of a population \parencite{edmonds2004mutations,klopfstein2005fate}. Because surfing can lead to the increase or even fixation of a given allele (be it neutral or not), it is easily mistaken for the product of selective forces. Yet unlike selection, surfing can also cause deleterious variants to increase and result in severe fitness loss at expanding fronts, termed expansion load \parencite{peischl2013accumulation}. 
This demographic process alters the actions of natural selection and genetic drift within the genome and has potentially large effects on population fitness, emphasizing its importance as a non-adaptive force in evolution.

\section*{Molecular-level processes}
\subsection*{Linked selection} \label{linkage}
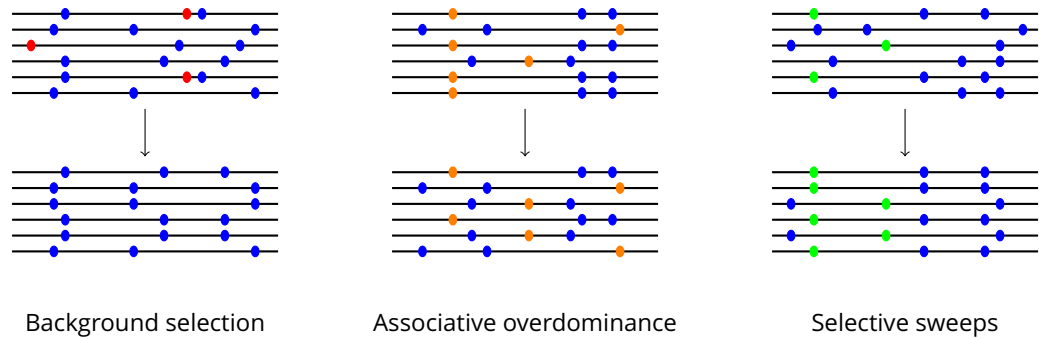
\begin{figure}[ht]
\begin{center}
\begin{tikzpicture}[xscale=0.5, yscale =0.7]

\draw[thick] (0,4.5) -- (7,4.5) ;
\draw[thick] (0,4.2) -- (7,4.2) ;
\draw[thick] (0,3.9) -- (7,3.9) ;
\draw[thick] (0,3.6) -- (7,3.6) ;
\draw[thick] (0,3.3) -- (7,3.3) ;
\draw[thick] (0,3) -- (7,3) ;

\draw[thick] (10,4.5) -- (17,4.5) ;
\draw[thick] (10,4.2) -- (17,4.2) ;
\draw[thick] (10,3.9) -- (17,3.9) ;
\draw[thick] (10,3.6) -- (17,3.6) ;
\draw[thick] (10,3.3) -- (17,3.3) ;
\draw[thick] (10,3) -- (17,3) ;

\draw[thick] (20,4.5) -- (27,4.5) ;
\draw[thick] (20,4.2) -- (27,4.2) ;
\draw[thick] (20,3.9) -- (27,3.9) ;
\draw[thick] (20,3.6) -- (27,3.6) ;
\draw[thick] (20,3.3) -- (27,3.3) ;
\draw[thick] (20,3) -- (27,3) ;

\draw[thick] (0,1.5) -- (7,1.5) ;
\draw[thick] (0,1.2) -- (7,1.2) ;
\draw[thick] (0,.9) -- (7,.9) ;
\draw[thick] (0,.6) -- (7,.6) ;
\draw[thick] (0,.3) -- (7,.3) ;
\draw[thick] (0,0) -- (7,0) ;

\draw[thick] (10,1.5) -- (17,1.5) ;
\draw[thick] (10,1.2) -- (17,1.2) ;
\draw[thick] (10,.9) -- (17,.9) ;
\draw[thick] (10,.6) -- (17,.6) ;
\draw[thick] (10,.3) -- (17,.3) ;
\draw[thick] (10,0) -- (17,0) ;

\draw[thick] (20,1.5) -- (27,1.5) ;
\draw[thick] (20,1.2) -- (27,1.2) ;
\draw[thick] (20,.9) -- (27,.9) ;
\draw[thick] (20,.6) -- (27,.6) ;
\draw[thick] (20,.3) -- (27,.3) ;
\draw[thick] (20,0) -- (27,0) ;

\draw [->] (3.5,2.7) -- (3.5,1.8);
\draw [->]  (13.5,2.7) -- (13.5,1.8);
\draw [->] (23.5,2.7) -- (23.5,1.8);
\node [below, fill= white] at (3.5,-1) {Background selection};
\node [below, fill= white] at (13.5,-1) {Associative overdominance};
\node [below, fill= white] at (23.5,-1) {Selective sweeps};

\draw [fill, blue] (1.4,4.5) circle [radius=0.1];
\draw [fill, red] (4.6,4.5) circle [radius=0.1];
\draw [fill, blue] (5,4.5) circle [radius=0.1];
\draw [fill, blue] (1.4,3.3) circle [radius=0.1];
\draw [fill, red] (4.6,3.3) circle [radius=0.1];
\draw [fill, blue] (5,3.3) circle [radius=0.1];
\draw [fill, blue] (1.4,3.6) circle [radius=0.1];
\draw [fill, blue] (4,3.6) circle [radius=0.1];
\draw [fill, blue] (5.6,3.6) circle [radius=0.1];
\draw [fill, red] (0.5,3.9) circle [radius=0.1];
\draw [fill, blue] (4.4,3.9) circle [radius=0.1];
\draw [fill, blue] (6,3.9) circle [radius=0.1];
\draw [fill, blue] (1.1,4.2) circle [radius=0.1];
\draw [fill, blue] (3.2,4.2) circle [radius=0.1];
\draw [fill, blue] (6.4,4.2) circle [radius=0.1];
\draw [fill, blue] (1.1,3) circle [radius=0.1];
\draw [fill, blue] (3.2,3) circle [radius=0.1];
\draw [fill, blue] (6.4,3) circle [radius=0.1];

\draw [fill, blue] (1.4,0.3) circle [radius=0.1];
\draw [fill, blue] (4,0.3) circle [radius=0.1];
\draw [fill, blue] (5.6,0.3) circle [radius=0.1];
\draw [fill, blue] (1.4,1.5) circle [radius=0.1];
\draw [fill, blue] (4,1.5) circle [radius=0.1];
\draw [fill, blue] (5.6,1.5) circle [radius=0.1];
\draw [fill, blue] (1.4,0.6) circle [radius=0.1];
\draw [fill, blue] (4,0.6) circle [radius=0.1];
\draw [fill, blue] (5.6,0.6) circle [radius=0.1];
\draw [fill, blue] (1.1,0.9) circle [radius=0.1];
\draw [fill, blue] (3.2,0.9) circle [radius=0.1];
\draw [fill, blue] (6.4,0.9) circle [radius=0.1];
\draw [fill, blue] (1.1,1.2) circle [radius=0.1];
\draw [fill, blue] (3.2,1.2) circle [radius=0.1];
\draw [fill, blue] (6.4,1.2) circle [radius=0.1];
\draw [fill, blue] (1.1,0) circle [radius=0.1];
\draw [fill, blue] (3.2,0) circle [radius=0.1];
\draw [fill, blue] (6.4,0) circle [radius=0.1];

\draw [fill, green] (21.1,4.5) circle [radius=0.1];
\draw [fill, blue] (24,4.5) circle [radius=0.1];
\draw [fill, blue] (25.6,4.5) circle [radius=0.1];
\draw [fill, green] (21.1,3.3) circle [radius=0.1];
\draw [fill, blue] (24,3.3) circle [radius=0.1];
\draw [fill, blue] (25.6,3.3) circle [radius=0.1];
\draw [fill, blue] (21.6,3.6) circle [radius=0.1];
\draw [fill, blue] (25,3.6) circle [radius=0.1];
\draw [fill, blue] (26,3.6) circle [radius=0.1];
\draw [fill, blue] (21.6,3) circle [radius=0.1];
\draw [fill, blue] (25,3) circle [radius=0.1];
\draw [fill, blue] (26,3) circle [radius=0.1];
\draw [fill, blue] (22.5,4.2) circle [radius=0.1];
\draw [fill, blue] (21.2,4.2) circle [radius=0.1];
\draw [fill, blue] (26.6,4.2) circle [radius=0.1];
\draw [fill, blue] (20.5,3.9) circle [radius=0.1];
\draw [fill, blue] (26,3.9) circle [radius=0.1];
\draw [fill, green] (23,3.9) circle [radius=0.1];

\draw [fill, green] (21.1,0) circle [radius=0.1];
\draw [fill, blue] (24,0) circle [radius=0.1];
\draw [fill, blue] (25.6,0) circle [radius=0.1];
\draw [fill, green] (21.1,1.2) circle [radius=0.1];
\draw [fill, blue] (24,1.2) circle [radius=0.1];
\draw [fill, blue] (25.6,1.2) circle [radius=0.1];
\draw [fill, green] (21.1,0.6) circle [radius=0.1];
\draw [fill, blue] (24,0.6) circle [radius=0.1];
\draw [fill, blue] (25.6,0.6) circle [radius=0.1];
\draw [fill, green] (21.1,1.5) circle [radius=0.1];
\draw [fill, blue] (24,1.5) circle [radius=0.1];
\draw [fill, blue] (25.6,1.5) circle [radius=0.1];
\draw [fill, blue] (20.5,.3) circle [radius=0.1];
\draw [fill, blue] (26,.3) circle [radius=0.1];
\draw [fill, green] (23,.3) circle [radius=0.1];
\draw [fill, blue] (20.5,.9) circle [radius=0.1];
\draw [fill, blue] (26,.9) circle [radius=0.1];
\draw [fill, green] (23,.9) circle [radius=0.1];

\draw [fill, blue] (12.1,3.6) circle [radius=0.1];
\draw [fill, blue] (14.7,3.6) circle [radius=0.1];
\draw [fill, orange] (13.6,3.6) circle [radius=0.1];
\draw [fill, orange] (11.6,3.3) circle [radius=0.1];
\draw [fill, blue] (15,3.3) circle [radius=0.1];
\draw [fill, blue] (15.8,3.3) circle [radius=0.1];
\draw [fill, orange] (11.6,3) circle [radius=0.1];
\draw [fill, blue] (15,3) circle [radius=0.1];
\draw [fill, blue] (15.8,3) circle [radius=0.1];
\draw [fill, orange] (11.6,3.9) circle [radius=0.1];
\draw [fill, blue] (15,3.9) circle [radius=0.1];
\draw [fill, blue] (15.8,3.9) circle [radius=0.1];
\draw [fill, orange] (11.6,4.5) circle [radius=0.1];
\draw [fill, blue] (15,4.5) circle [radius=0.1];
\draw [fill, blue] (15.8,4.5) circle [radius=0.1];
\draw [fill, blue] (12.5,4.2) circle [radius=0.1];
\draw [fill, blue] (10.8,4.2) circle [radius=0.1];
\draw [fill, orange] (16,4.2) circle [radius=0.1];

\draw [fill, blue] (12.1,0.9) circle [radius=0.1];
\draw [fill, blue] (14.7,0.9) circle [radius=0.1];
\draw [fill, orange] (13.6,0.9) circle [radius=0.1];
\draw [fill, blue] (12.1,0.3) circle [radius=0.1];
\draw [fill, blue] (14.7,0.3) circle [radius=0.1];
\draw [fill, orange] (13.6,0.3) circle [radius=0.1];
\draw [fill, orange] (11.6,0.6) circle [radius=0.1];
\draw [fill, blue] (15,0.6) circle [radius=0.1];
\draw [fill, blue] (15.8,0.6) circle [radius=0.1];
\draw [fill, orange] (11.6,1.5) circle [radius=0.1];
\draw [fill, blue] (15,1.5) circle [radius=0.1];
\draw [fill, blue] (15.8,1.5) circle [radius=0.1];

\draw [fill, blue] (12.5,1.2) circle [radius=0.1];
\draw [fill, blue] (10.8,1.2) circle [radius=0.1];
\draw [fill, orange] (16,1.2) circle [radius=0.1];
\draw [fill, blue] (12.5,0) circle [radius=0.1];
\draw [fill, blue] (10.8,0) circle [radius=0.1];
\draw [fill, orange] (16,0) circle [radius=0.1];

\end{tikzpicture} 
 \caption{\label{figure:hitchhiking} Genetic linkage can drastically change the frequency of neutral alleles in a population, falling into three categories depending on the manner of selection towards a focal site. For each form of selection shown, the top row shows the haplotypes of a non-recombining region in the initial population. The bottom row shows the resultant haplotypes after an episode of selection. For illustrative purpose, each haplotype contains 3 derived alleles (circles): neutral ones are in blue, beneficial in green, and deleterious alleles are in red or orange to indicate dominant or recessive, respectively. For background selection (left), neutral diversity is reduced due to negative selection on nearby linked deleterious alleles and homozygosity increases at the population level. Associative overdominance (center) prevents combinations of homozygous neutral alleles from accumulating in the population in regions of low recombination. At the beginning, each haplotype contains one deleterious recessive allele (orange circle) and loci carrying such alleles are genetically linked as the region is non-recombining. Selection favors combinations of heterozygous deleterious alleles as they are recessive. Neutral heterozygosity is favored at the population level and diversity increases. Selective sweeps (right) reduce diversity and increase allele frequency through the hitchhiking of neutral variants that are linked to beneficial mutations under positive selection. Figure inspired from Alves et al. \parencite*{Alves2012}.}
\end{center}
\end{figure}

The fact that recombination breaks apart combinations of alleles at an increasing probability with greater distance along the genome results in many sites being physically linked and evolving in a non-independent manner. The background where a new mutation occurs therefore influences that variant's probability of fixation, as any more strongly-selected target sites nearby will influence that linked site, as first pointed out by Fisher \parencite*{Fisher1930} and Muller \parencite*{Muller1932} (Figure \ref{figure:hitchhiking}). A extended review on this topic was written by Gordo and Charlesworth \parencite*{gordo}.

Neutral sites in a background with one or more sites under negative selection will have a lower probability of fixation than unlinked neutral sites. This is due to background selection (BGS), where negative selection against a variant reduces the frequency of nearby neutral variants \parencite{charlesworth1993}. Sites subject to BGS fall in the category of non-adaptive evolution because these variants are not directly selected against nor are they evolving neutrally since selection indirectly impacts them. These linked sites evolve in a non-neutral fashion, so even if phenotypically and adaptively they confer no change in phenotype, they must not be considered neutral for inferential purposes, a point which is widely recognized in the field.

Similarly, the occurrence of a mutation conferring a fitness benefit can also result in the reduction of genetic diversity through a selective sweep (Figure \ref{figure:hitchhiking}). When selection increases the frequency of a beneficial allele in the population, nearby neutral variants likewise increase in frequency, hitchhiking along to fixation with the beneficial variant. Whether selection acts on a single novel variant \parencite[hard sweeps;][]{messer} or on standing genetic variation \parencite[soft sweeps;][]{softsweep,softsweeps2,softsweep3} can influence the extent of the impact on allele frequency change for linked neutral sites. Several population level processes may even be contributors to instances where standing genetic variation results in a soft sweep, for example if existing diversity shifts to become beneficial, perhaps due to environmental change.

Finally, genetic linkage can also lead to an increase in genetic diversity when neutral sites fall near a partially deleterious recessive allele or near an allele under balancing selection. In the presence of partially recessive deleterious alleles, this increase in diversity is termed associative overdominance \parencite[AOD;][]{ohtakimura1970,zhao2016resolving,gilbert2020transition,becher2020patterns}, and is limited to regions of low recombination (Figure \ref{figure:hitchhiking}). In contrast, sites linked to those under classical balancing selection should increase in diversity across regions spanning the range of possible recombination rates. 

Most methods to detect selection on linked polymorphisms are based on the site frequency spectrum (SFS), such as Tajima's D which depends on the pairwise nucleotide diversity and the number of segregating sites \parencite{tajima} and other methods which are based on the haplotype structure \parencite{hudson1995,sabeti2002,voight2006,sabeti2007}. Methods based on haplotype structure are most effective to detect recent episodes of hitchhiking \parencite{garud2}.

\subsection*{Hypermutability of CpG sites and mutation rate variation}
CpG sites in which a cytosine and a guanine appear consecutively, can experience high levels of mutational pressure. Cytosines at CpG sites are one of the preferential targets of methylation in vertebrates and some other species. Methylated cytosines spontaneously deaminate to thymines, leading to an increase in the frequency of TpG sites within the genome and a relative deficit of CpG \parencite[reviewed in][]{Hodginkson}, potentially leaving a signature indicative of selection.

A recent paper by Laurin-Lemay et al. \parencite*{cpgnatsel} found that a large proportion of mammalian codon usage, such as the preferential usage of the GCC Alanine codon compared to its synonymous GCG in humans, can be explained by the hypermutability of CpG sites, even though this is often unaccounted for in codon substitution models. The authors advocate for evaluating the impact of such model violations on statistical tests in phylogenetic analyses. Interestingly, CpG hypermutability is also an underappreciated process in the field of population genomics where the favored strategy has been to filter out hypermutable sites before performing evolutionary inferences \parencite{pouyet2018}. The hypermutability of these sites and the subsequent bias to certain alleles has been shown to shift site frequency spectra in ways that might interfere with population genetic inferences that are based on the SFS \parencite{harpak}.

\subsection*{Segregation distortion}\label{section:segregationdistortion}
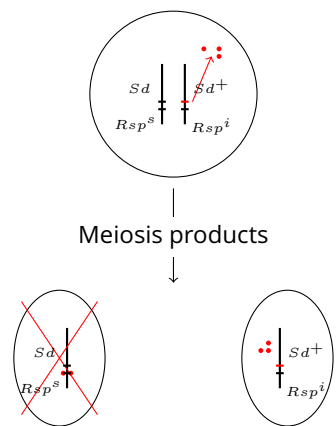
\begin{figure}[ht]
\begin{center}
\begin{tikzpicture}[scale=0.5]
\draw (5,7) circle (2.2);
\draw (2,0) ellipse (1.2cm and 1.8cm);
\draw (8,0) ellipse (1.2cm and 1.8cm);
\draw [->] (5,4.5) -- (5,2);
\node [below, fill=white] at (5,3.75) {Meiosis products};
\draw[thick,fill] (4.7,6.2) -- (4.7,7.8) ;
\draw[thick,fill] (5.3,6.2) -- (5.3,7.8) ;
\draw[thick,fill] (2.2,-0.8) -- (2.2,0.8) ;
\draw[thick,fill] (7.8,-0.8) -- (7.8,0.8) ;

\draw[thick] (4.6,6.6) -- (4.8,6.6) ;
\node [left, below] at (4,6.6) {\tiny $Rsp^s$};
\draw[thick] (4.6,6.8) -- (4.8,6.8) ;
\node [left, above] at (4.1,6.8) {\tiny $Sd$};

\draw[thick] (5.2,6.6) -- (5.4,6.6) ;
\node [right, below] at (6,6.6) {\tiny $Rsp^i$};
\draw[->, red] (5.5,6.8) -- (6,8) ;
\draw [fill, red] (5.8,8.2) circle [radius=0.05];
\draw [fill, red] (6.2,8) circle [radius=0.05];
\draw [fill, red] (6.2,8.2) circle [radius=0.05];
\draw[thick,red] (5.2,6.8) -- (5.4,6.8) ;
\node [right, above] at (6,6.8) {\tiny $Sd^+$};

\draw [fill, red] (2.1,-0.4) circle [radius=0.05];
\draw [fill, red] (2.3,-0.4) circle [radius=0.05];
\draw [fill, red] (1,-1.5) -- (3,1.5) ;
\draw [fill, red] (3,-1.5) -- (1,1.5) ;
\draw[thick] (2.1,-0.4) -- (2.3,-0.4) ;
\node [left, below] at (1.5,-0.4) {\tiny $Rsp^s$};
\draw[thick] (2.1,-0.2) -- (2.3,-0.2) ;
\node [left, above] at (1.6,-0.2) {\tiny $Sd$};

\draw[thick] (7.7,-0.4) -- (7.9,-0.4) ;
\node [right, below] at (8.5,-0.4) {\tiny $Rsp^i$};
\draw [fill, red] (7.3,0.2) circle [radius=0.05];
\draw [fill, red] (7.5,0.2) circle [radius=0.05];
\draw [fill, red] (7.5,0.4) circle [radius=0.05];
\draw[thick,red] (7.7,-0.2) -- (7.9,-0.2) ;
\node [right, above] at (8.5,-0.2) {\tiny $Sd^+$};

\end{tikzpicture} 
\caption{\label{figure:segregationdistortion} Figure simplified from \citet{meioticdrive}. An example of segregation distortion of meiotic gene drivers with the ``killer-target" strategy is shown for the SD gene complex in \textit{Drosophila melanogaster}. This complex involves the $Sd$ and $Rsp$ loci. The ``target" locus $Rsp$ harbours two alleles: the $Rsp^s$ and $Rsp^i$ that are respectively sensitive and insensitive to the meiotic driver. $Sd$ produces a ``killer" element (red dots, a protein in this example) which interferes with $Rsp^s$ and kills the meiotic products that inherit $Rsp^s$.} 
\end{center} 
\end{figure}

Meiotic gene drivers manipulate the transmission process during meiosis to their own advantage, leading to their over-representation in gametes despite the lack of advantage to the carrier. As segregation distortion encompasses a wide variety of mechanisms, we provide three such examples. The first one is the segregation distorter gene complex \parencite[Figure \ref{figure:segregationdistortion};][]{Larracuente}, present at low frequency in all natural populations of \textit{Drosophila melanogaster}. This complex involves the $Sd$ locus and its target responder locus ($Rsp$). There is variable transmission advantage between the different $Sd$ alleles, and one of these alleles recently swept to fixation in Africa causing strong linkage disequilibrium and loss of genetic diversity \parencite{pregraves}. A second example is \textit{Wolbachia}, where a maternally inherited bacteria of arthropods manipulates host reproduction. \textit{Wolbachia} is known to be a selfish element favoring its own propagation through, for instance, the inability of infected males to successfully reproduce with uninfected females \parencite{cytoIncompatibility}. At the level of the host, the presence of \textit{Wolbachia} is also associated to segregation distortion: the maternal inheritance induces genetic linkage on host mitochondria and mitochondria in infected females are over-represented in the next generation. This effect reduces the effective population size and the efficacy of selection in mitochondria and could drive fixation of mitochondrial haplotypes \parencite{HurstJiggings_wolbachia, cariou}. 
A third example comes from centromere evolution, where asymmetry at female meiosis causes only one of the four products of meiosis to become the oocyte nucleus and can lead to a kind of segregation distortion \parencite{Henikoff}. Centromeres have a central role in preventing aneuploidy by facilitating the assembly of several components required for chromosome separation \parencite[see][for a review]{centromererole}. This centromere drive model includes proteins such as Cid and highly repetitive satellite sequences that bind to microtubules during meiosis I \parencite{Henikoff}. Centromeres which preferentially transmit to the oocyte nucleus can rapidly drive to fixation even with a slight advantage at each meiosis \parencite{Henikoff}.

Meiotic drivers are predicted to be evolutionarily 
labile by favoring their fixation in the population even though they are detrimental for their carriers \parencite{LINDHOLM2016}. Identifying loci under selection in these cases is anything but straightforward, as the signal of meiotic drive might be easily confounded with selective sweeps and positive selection \parencite{pregraves}.

\subsection*{Mobility of genetic elements}
Detecting the signature of selection is often restricted to SNPs even though transposable elements (TE) and other structural genomic changes such as inversions are likely also prevalent biases for adaptive evolution. 
Until further studies improve our understanding of mobile genetic elements and our ability to identify them, it is certainly possible that these regions of the genome may play a role in changing diversity and biasing or interrupting our ability to infer selection or neutral demographic parameters, e.g. in inverted regions that can no longer recombine, deleterious variation can become masked or beneficial variation may be maintained in tight linkage. 

TEs are widespread across the tree of life and in some species can represent a major fraction of the genome \parencite{wicker,dekoning}. TEs are associated with the creation of new mutations and changes in recombination patterns  \parencite[e.g.][]{bartolome}. McClintock \parencite*{McClintock} first discovered that mobile elements were associated with phenotypic changes in maize. 
However, these sorts of structural genomic changes are often disregarded in favor of SNPs when inferring sources of adaptive evolution mainly because of methodological limitations \parencite{villanueva2017}. First, TE families are difficult to identify as they are repetitive elements spread throughout the genome with limited descriptive features such as target site duplications or terminal repeats and transposases \parencite{xiong}. They are also commonly associated with genetic load as they can lead to diseases if inserted into genes \parencite{chenTE}. For instance, the Alu family of insertions are associated to haemophilia or breast cancer in humans \parencite{aluhuman}. Second, because repetitive elements are enriched for TEs, it is difficult to assemble these genomic regions \parencite{yannbourgeois}. For the past decade, new techniques have emerged to infer selection on TE insertions and are divided into two main classes: SFS-based or haplotype-based methods \parencite{villanueva2017}. It is clear that as our understanding of TE dynamics improves, such knowledge may greatly contribute to our understanding of selection acting at the genomic level and modes of adaptive evolution.  

\subsection*{Gene conversion}\label{bgc}
Meiotic recombination reshuffles the genetic material of parents to produce a new set of genetic material in offspring. During recombination, homologous gene conversion can result from the conversion of an acceptor locus at heterozygous sites in donor sequences. Biased gene conversion (BGC),occurring at locations where recombination breaks the DNA strand, makes the probability of transmitting one of the two alleles larger than the probability of losing it. BGC is comprised of two main mechanisms: double-strand-break-driven \parencite[dBGC;][]{Myers, prdm9review} and GC-driven \parencite[gBGC;][]{duret2009a,lesecqueBGC}, each of which have different mechanistic origins and consequences. To our knowledge dBGC is not expected to be confounded with selection and will not be discussed further herein. On the contrary, gBGC is often responsible for false positives in inferences of selection \parencite{Galtieretal2009,pmid20643747}.

\begin{figure}[ht]
\begin{center}
\begin{tikzpicture}[scale=2]

\draw [ red,ultra thick] (0,4) -- (1.7,4);
\draw [ blue,ultra thick] (1.7,4) -- (2.3,4);
\draw [->, blue,ultra thick] (2.3,4) -- (3,4);
\draw [<-, red,ultra thick] (0,3.8) -- (1.1,3.8);
\draw [blue,ultra thick] (1.1,3.8) -- (3,3.8);
\draw [ blue,ultra thick] (0,3.5) -- (2.3,3.5);
\draw [->, red,ultra thick] (2.3,3.5) -- (3,3.5);
\draw [ red,ultra thick] (1.7,3.3) -- (3,3.3);
\draw [  blue,ultra thick] (1.7,3.3) -- (1.1,3.3); 
\draw [<-, blue,ultra thick] (0,3.3) -- (1.1,3.3);
\draw [draw=black, fill=gray, opacity=0.2] (1.1,3.7) -- (1.7,3.7) -- (1.7,4.1) -- (1.1,4.1) -- cycle;
\draw [draw=black, fill=gray, opacity=0.2] (1.7,3.2) -- (2.3,3.2) -- (2.3,3.6) -- (1.7,3.6) -- cycle;
\node[above] at (2,3.35) {T};
\node[below] at (2,3.45) {C};
\draw [ blue,ultra thick] (0.9,2) -- (1.5,2);
\draw [ red,ultra thick] (0.9,1.8) -- (1.5,1.8);
\node[above] at (1.2,1.85) {G};
\node[below] at (1.2,1.95) {C};
\draw [ blue,ultra thick] (2.5,2) -- (3.1,2);
\draw [ red,ultra thick] (2.5,1.8) -- (3.1,1.8);
\node[above] at (2.8,1.85) {T};
\node[below] at (2.8,1.95) {A};
\draw [line width=1mm, ->, gray] (1.9,3) -- (1.4,2.5);
\draw [line width=0.3mm, ->, gray ] (2.1,3) -- (2.6,2.5);
\end{tikzpicture}
\caption{\label{figure:bgc} Figure adapted from Glémin et al. \parencite*{GLEMIN2014}. GC-biased gene conversion (gBGC) occurs after the formation of hetero-duplexes during a meiotic recombination event. Heteroduplexes can result from crossing-over (shown here) as well as from non-crossover events. In the presence of gBGC, mismatches are repaired more than half of the time in favor of guanine and cytosine rather than adenine and thymine. This bias towards enrichment of guanine and cytosine is of few percent in yeast or humans, leading to the accumulation of GC content within genomes over time.}
\end{center}
\end{figure}
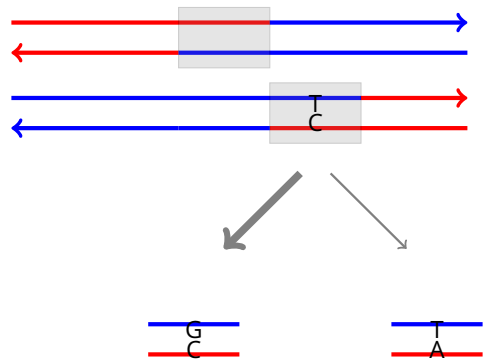
gBGC is a transmission bias in favor of G/C over A/T alleles when a mismatch is repaired after a meiotic recombination event. This leads to the increase of GC content in regions of high recombination over evolutionary time \parencite{duret2009a} (Figure \ref{figure:bgc}). Evidence for gBGC has been shown in many organisms \parencite{duret2009a,lesecqueBGC,webster2006, pmid25659072} and has strong consequences on genomic architecture, ranging from global GC-enrichment of genomes to variation in codon usage between genes \parencite[][]{pouyet2017}, and can be confounded with translational selection \parencite{gingold}. 
When G/C alleles are beneficial as compared to A/T, gBGC amplifies the speed of fixation of a beneficial allele in the population, while in other cases when G/C alleles are slightly deleterious, gBGC counteracts the effect of selection (if the transmission bias parameter, $b$, is stronger than the selection parameter, $s$. At the population level, gBGC shifts the site frequency spectrum towards the left for strong-to-weak polymorphisms (respectively to the right for weak-to-strong) mimicking the effect of natural selection in high recombination rate regions \parencite{pouyet2018,lachancegbgc,glemin2015}. One way to consider the consequences of gBGC is to contrast genetic diversity of a region between weak-to-strong and strong-to-weak mutations. Additionally, gBGC should not leave the same signature as linked selection because gBGC acts solely on its targets sites and does not affect the surrounding diversity.
There is no inherent selective bias driving the genomic changes resulting from gene conversion, nor can a biased process by definition be considered neutral. Instead, this process is best considered as non-adaptive evolution, where fitness is not impacted, but there is clearly a non-neutral change in allele frequencies over time.
As illustrated by Hurst \parencite*{Hurst2019}, segregation distortion and BGC are rarely considered simultaneously even though they act similarly to increase their transmission to the detriment of other alleles. Both BGC and segregation distortion along with structural genomic changes are common in nature, and including these processes in studies of adaptive evolution will allow us to properly identify targets of positive selection \parencite{villanueva2017}.

\section*{Conclusion}
\label{Ccl}
In the modern genomic era, emerging data will hopefully allow us to build a complete picture of the relevant genomic, demographic, and environmental scenarios where different evolutionary processes are expected to dominate changes in molecular diversity over time. Identifying a variant as subject to natural selection is difficult since the selective environment of an allele is a combination of its own innate properties impacting the genome, along with epistatic effects due to its genomic environment, as well as its demographic situation (how efficient selection is in the population where this individual exists), and lastly its (a)biotic environment (e.g. stressful environments for the organism harboring this variant, or the frequency of conspecific phenotypes). Understanding the processes that may bias our inferences of sites under selection is paramount to better understanding the evolutionary forces leading to genomic change. To a large extent these biases result from the difficulty or inability to distinguish non-adaptive sites from sites under direct selection. As discussed, this is largely due to the demographic processes in a population’s past that make otherwise neutral sites mimic selection and fall into the category of non-adaptive, as well as due to the molecular processes that change neutral allele frequencies in biased ways and can even counteract the effect of selection on selected sites (gBGC, for instance). 
There is an implied distinction worth explicitly stating: even if the majority of sites within the genome may be neutral in terms of their selection coefficient, it is very likely the case that the majority of the genome evolves due to the impact of selective forces, even if that targets few specific sites, due to the degree of linkage within the genome. This still does not, however, discount the fact that non-adaptive evolutionary processes have an important impact on genomic change.
Incorporating all of this information in future studies is a tall task, particularly since empirical study of biology is further complicated by changing environments and demographics that are not always apparent to observers, nor always sufficiently sampled. We hope that this perspective has highlighted the importance of recognizing and distinguishing the complex interactions of selective, non-adaptive, and neutral processes acting within and among genomes and serves to move the field of evolutionary genomics forward in understanding the drivers of molecular diversity.

\paragraph{Conflict of interest disclosure}
The authors of this article declare that they have no financial conflict of interest with the content of this article.
\section*{Acknowledgements}
We would like to thank Benoit Nabholz, an anonymous reviewer, and Guillaume Achaz for critical feedback, as well as Tyler Kent, Stephan Peischl, and Marie Cariou for their input on earlier versions, all of which greatly improved this manuscript. \\ Version 4 of this preprint has been peer-reviewed and recommended by Peer Community In Evol Biol (https://doi.org/10.24072/pci.evolbiol.100103)


\printbibliography[notcategory=ignore]

\end{document}